\documentclass[twocolumn,showpacs,preprintnumbers,amsmath,amssymb,prb,superscriptaddress]{revtex4}
\usepackage{epsfig}
\usepackage{graphicx}
\usepackage{dcolumn}
\usepackage{bm}
\usepackage{color} 

\begin{document}

\title{
Momentum-dependent hybridization gap and dispersive in-gap state 
of The Kondo Semiconductor SmB$_6$
}
\author{Hidetoshi Miyazaki}
 \altaffiliation[Present address: ]{Center for Fostering Young and Innovative Researchers, Nagoya Institute of Technology, Nagoya 466-8555, Japan}
\affiliation{UVSOR Facility, Institute for Molecular Science, Okazaki 444-8585, Japan}
\author{Tetsuya Hajiri}
\affiliation{Graduate School of Engineering, Nagoya University, Nagoya 464-8603, Japan}
\affiliation{UVSOR Facility, Institute for Molecular Science, Okazaki 444-8585, Japan}
\author{Takahiro Ito}
\affiliation{Graduate School of Engineering, Nagoya University, Nagoya 464-8603, Japan}
\author{Satoru Kunii}
\affiliation{Department of Physics, Tohoku University, Sendai 980-8578, Japan}
\author{Shin-ichi Kimura}
 \altaffiliation[Electronic address: ]{kimura@ims.ac.jp}
\affiliation{UVSOR Facility, Institute for Molecular Science, Okazaki 444-8585, Japan}
\affiliation{School of Physical Sciences, The Graduate University for Advanced Studies (SOKENDAI), Okazaki 444-8585, Japan}
\date{\today}
\begin{abstract}
We report the temperature-dependent three-dimensional angle-resolved photoemission spectra of the Kondo semiconductor SmB$_6$.
We found a difference in the temperature dependence of the peaks at the X and $\Gamma$ points, due to hybridization between the Sm~$5d$ conduction band and the nearly localized Sm~$4f$ state.
The peak intensity at the X point has the same temperature dependence as the valence transition below 120~K, while that at the $\Gamma$ point is consistent with the magnetic excitation at $Q=(0.5,0.5,0.5)$ below 30~K.
This suggests that the hybridization with the valence transition mainly occurs near the X point, and the initial state of the magnetic excitation is located near the $\Gamma$ point.
\end{abstract}
\maketitle
%
%
%
Materials with strong electron correlation have exotic physical properties that cannot be predicted from first-principle band calculations.
One example may be seen in a semiconductor with a very small energy gap, which appears in rare-earth compounds such as the Kondo semiconductor/insulator (KI)~\cite{Aeppli1992}.
At high temperatures, KI behaves like a dense Kondo metal, while an energy gap with activation energy of several 10~meV appears at low temperature.
The energy gap is believed to originate from hybridization between the nearly localized $4f$ state near the Fermi level ($E_F$) and the conduction band ($c$-$f$ hybridization).

Numerous studies have investigated the energy gap of KI, using optical conductivity~\cite{Travaglini1984,Kimura1994}, point contact spectroscopy~\cite{Flachbart2001}, angle-integrated photoemission spectroscopy~\cite{Nozawa2002,Souma2002}, and other methods.
However, the momentum dependence of the $c$-$f$ hybridization gap, as well as the relation of the electronic structure to other physical properties, has yet to be studied.
Because the $c$-$f$ hybridization occurs at a specific momentum vector, the most direct method of observing the band dispersion of the $c$-$f$ hybridization gap is three-dimensional angle-resolved photoemission spectroscopy (3D-ARPES) using a tunable photon source from synchrotron radiation.
Thus, we applied the 3D-ARPES method to observe the $c$-$f$ hybridization gap creation of a typical KI, SmB$_6$.

SmB$_6$ is a valence-fluctuation material in between Sm$^{2+}$($4f^6$) and Sm$^{3+}$($4f^5$) ions~\cite{Cohen1970}.
The electrical resistivity ($\rho$) decreases on cooling, like a metal, above the temperature of 100~K, but then reveals a semiconductor-like character with activation energy of 15~meV~\cite{Tanaka1980}.
There are two characteristic temperatures on SmB$_6$; one is valence transition below 120~K, and the other is magnetic excitation below 30~K.
The mean valence changes from 2.57 at 120~K to 2.50 at 40~K on cooling~\cite{Mizumaki2009}.
Coincidentally, the lattice constant, which normally shrinks above 120~K on cooling, anomalously expands from 120~K to a few tens~K indicating the valence change from Sm$^{3+}$ to Sm$^{2+}$~\cite{Kasuya1981}.
On the other hand, the magnetic excitation at the scattering vector of $Q=(0.5,0.5,0.5)$, observed by inelastic neutron scattering (INS), rapidly increases below 30~K~\cite{Alekseev1995}.
Then the mean valence slightly recovers from 2.50 to 2.52 below 30~K, and the lattice constant shrinks again.
The reason for the different temperature dependence between the magnetic excitation and the valence transition has yet to be determined.
At temperatures lower than 10~K, another gap (in-gap state) has been noted at about 4~meV through the observations of optical conductivity~\cite{Kimura1994,Gorshunov1999}, point-contact spectroscopy~\cite{Flachbart2001}, and angle-integrated photoemission spectroscopy~\cite{Nozawa2002}.
Below 3~K, $\rho$ becomes saturated and has a residual resistivity of several $\Omega\cdot$cm~\cite{Cooley1995}.
Recently, the residual resistivity has been suggested to originate from metallic behavior at the edge state on the surface [topological KI (TKI)]~\cite{Dzero2010} and other reasons, but the relation of the electronic state to the in-gap state, has never been investigated experimentally.

In this paper, we report the temperature dependence of the dispersion curve of the hybridization state using temperature-dependent 3D-ARPES, in order to determine the electronic structure and the reason for the different temperature dependence of the valence transition and magnetic excitation.
We found that the hybridization band with a peak at the binding energy ($E_B$) of 15~meV near the X point gradually appears on cooling from 150 to 40~K, which has the same temperature dependence as the valence transition.
At the $\Gamma$ point, on the other hand, the peak at $E_B\sim$20~meV has the same temperature dependence as the magnetic excitation at $Q=(0.5,0.5,0.5)$, which differs from the 15-meV peak at the X point.
This suggests that the magnetic excitation originates from the hybridization band at the $\Gamma$ point.

%
%
\begin{figure}[t]
\begin{center}
\includegraphics[width=0.45\textwidth]{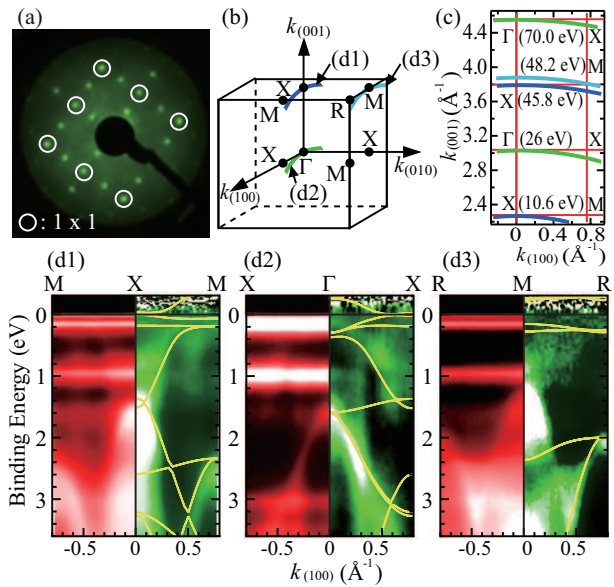}
\end{center}
\caption{
(Color online)
(a) Low energy electron diffraction (LEED) pattern of the SmB$_6$ $(001)$ surface with the electron energy of 100~eV.
Spots marked by circles represent diffractions of $1\times1$; other spots represent $1\times2$ and higher order diffractions.
(b) The first Brillouin zone of SmB$_6$ and high-symmetry points.
(c) The photon energies corresponding to the $\Gamma$ and X points in normal emission geometry and the M point in grazing emission geometry.
(d1)--(d3) ARPES intensity mapping images and calculated band dispersions (solid lines) on the M--X ($h\nu$=45.8~eV), X--$\Gamma$ (70.0~eV), and R--M (48.2~eV) lines at 10~K.
The left-hand images in each panel are the ARPES intensity mappings, which emphasize flat bands, and the right-hand images are the same intensity mappings but divided by angle-integrated photoemission spectra, which emphasize highly dispersive bands.
}
\label{Wide}
\end{figure}
A high purity single crystal of SmB$_6$ was grown by the floating-zone method~\cite{Tanaka1975}.
The sample cut along the (001) plane was cleaned by argon sputtering and annealing at 1400~$^\circ$C, by using an infrared heating system 
under a vacuum of $10^{-8}$~Pa.
The low-energy electron diffraction (LEED) image of the clean surface is shown in Fig.~\ref{Wide}(a).
The LEED image indicates that the SmB$_6$ (001) surface has not only $1\times1$ of the bulk but also a superlattice structure of $2\times1$ of the surface, which is similar to the previous result~\cite{Aono1978}.

Since the peak intensity of the $2\times1$ diffraction is about half of that of the $1\times1$ diffraction and the background is not higher than the previous result, the well-defined surface state was obtained.
However, the surface state is mostly boron-terminated B$_6$; i.e., the Sm atoms do not appear as much on the surface.
Therefore the surface state derived from Sm is expected to be suppressed.

Normal-emission and in-plane ARPES experiments in the vacuum-ultraviolet region were performed at beamlines 5U~\cite{Ito2007} and 7U~\cite{Kimura2010}, respectively, of the UVSOR-II storage ring, Institute for Molecular Science.
The photon energies corresponding to high symmetry $k_z$ points, as shown in Fig.~\ref{Wide}(b) were determined by using normal emission ARPES taken at BL5U (Fig.~\ref{Wide}c).
The inner potential was determined as 13.5~eV.
Using the obtained high symmetry points, the temperature dependence of high-resolution ARPES spectra at the $\Gamma$ and X points was measured at BL7U.
The total energy resolutions were about 50~meV at BL5U and 5~meV at BL7U, and the vacuum during the measurement was less than $5\times10^{-9}$~Pa.
LDA band structure calculation was performed by the full potential linearized augmented plane wave plus local orbital (LAPW+lo) method including spin-orbit coupling implemented in the {\sc Wien2k} code~\cite{WIEN2k}.

%
%
Figures~\ref{Wide}(d1)--\ref{Wide}(d3) are the ARPES images of the M--X, X--$\Gamma$, and R--M lines, taken at photon energies of 45.8, 70.0, and 48.2~eV, respectively, and the corresponding band calculation results.
The photon energies at the high symmetry points are consistent with a previous work~\cite{Denlinger2000}.
The flat bands at the binding energies ($E_B$) of 0, 0.2, 1, and 3~eV can be recognized as the multiplet structures of $^6H_{5/2}$, $^6H_{7/2}$, $^6F$, and $^6P$ of Sm$^{2+}$ final state, respectively.
There is another flat band at $E_B\sim$1.5~eV, which seems to originate from the surface state of Sm~$4f$, but the intensity is not great, because the surface is terminated by boron atoms.
The highly dispersive valence bands at $E_B\geq$1.4~eV and the conduction band at $E_B\leq$1.5~eV at the X point originate from the $sp$ covalent state of the B$_6$ network and the Sm~$5d$ state, respectively.
These higher-$E_B$ bands are in good agreement with the band structure calculation with a hole-like band appearing at $E_B\geq$1~eV owing to $k_z$-broadening.
The intensities of the Sm$^{2+}$ final state multiplet $^6H$ and $^6F$ are greater near the X point than those at the M point.
This implies that the hybridization occurs near the X point.
The band calculation in the right-hand images of Figs.~\ref{Wide}(d1) and \ref{Wide}(d2) also indicated that the hybridization gap opens near the X point.
It should be noted that, however, the band calculation has much overstated Sm $4f$ band widths and hybridization effects, the failure to describe the mixed valency, and lack of final-state multiplet structure.

\begin{figure}[t]
\begin{center}
\includegraphics[width=0.45\textwidth]{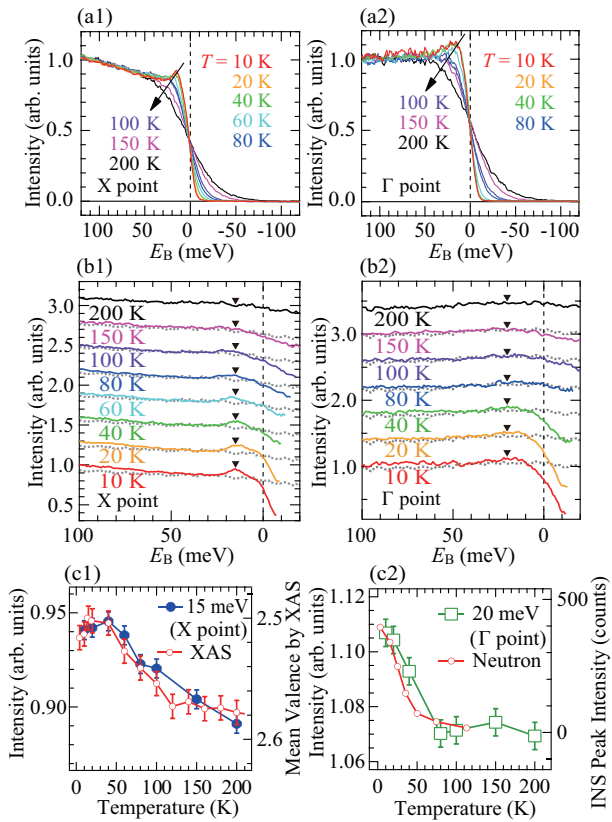}
\end{center}
\caption{
(Color online)
Temperature dependence of energy distribution curves (EDCs) at the X point ($h\nu$~=~10.6~eV, a1) and the $\Gamma$ point (26~eV, a2), 
and their spectra divided by the Fermi-Dirac distribution curve (b1, b2).
EDCs were normalized at the binding energy $E_B$ of 120~meV.
Successive curves in b1 and b2 are offset by 0.3 and 0.4, respectively, for clarity.
The dotted lines in (b1) and (b2) are the spectra at 200~K.
The temperature-dependent peak intensities relative to the intensity at 200~K of the 15-meV peak at the X point (c1) and of the 20-meV peak at the $\Gamma$ point (c2).
The mean valence number evaluated by the X-ray absorption spectroscopy (XAS)~\cite{Mizumaki2009} and the peak intensity of the inelastic neutron scattering (INS) at $Q=(0.5,0.5,0.5)$~\cite{Alekseev1995}, are also plotted.
}
\label{TdepGX}
\end{figure}
To investigate the $c$-$f$ hybridization gap formation, we measured the temperature dependence of the Sm~$5d$ character of the hybridization gap (peak) in energy distribution curves (EDCs) at the X and $\Gamma$ points, as shown in Fig.~\ref{TdepGX}.
Because of the hybridization between the Sm~$5d$ and $4f$ states, the $4f$ feature must appear in the Sm~$5d$ EDC.
Then we used low photon energies of 10.6~eV for the X point and 26~eV for the $\Gamma$ point, because the Sm~$4f$ cross section can be strongly suppressed by using lower energy photons below 30~eV~\cite{Yeh1985}.
The temperature dependence of EDCs obtained at the X and $\Gamma$ points is shown in Figs.~\ref{TdepGX}(a1) and \ref{TdepGX}(a2).
At 10~K, there are peaks at $E_B\sim$15~meV at the X point, and at 20~meV at the $\Gamma$ point.
These peaks have different temperature dependence.
Figures~\ref{TdepGX}(b1) and (b2) are the same spectra as in Figs.~\ref{TdepGX}(a1) and \ref{TdepGX}(a2), respectively, divided by the Fermi-Dirac distribution curve.
Clear energy gaps appear above $E_F$ at both the X and $\Gamma$ points.
Because the EDCs originate from the Sm~$5d$ states, the peaks at the gap edges are evidence of the hybridization with the Sm~$4f$ states.
At the X point, the peak becomes visible at 100~K and increases on cooling.
To clarify the relation of these peaks to other physical properties, the temperature dependence of the peak is plotted in Fig.~\ref{TdepGX}(c1).
The peak intensity gradually increases below 150~K.
The temperature dependence of the peak is very similar to the valence transition (2.57 at 120~K $\rightarrow$ 2.50 at 40~K) observed by using X-ray absorption spectroscopy at the Sm~$L_3$ edge, as shown in Fig.~\ref{TdepGX}(c1)~\cite{Mizumaki2009}.
Therefore the temperature dependence of the 15-meV peak at the X point indicates the change in the hybridization intensity, i.e., the hybridization gap opens near the X point.
This is roughly consistent with the LDA band calculation as shown in Fig.~\ref{Wide}, in which the hybridization between the Sm~$5d$ conduction band and $4f_{5/2}$ states appears near the X point, but the experimental $4f$ band width is much smaller than that in the calculation.
The discussion is done later.
On cooling, the hybridization becomes stable, and the mean valence shifts to divalent.

At the $\Gamma$ point, on the other hand, the temperature dependence of the 18-meV peak is not the same as the 15-meV peak at the X point.
The spectrum in Fig.~\ref{TdepGX}(b2) has no energy gap above 80~K.
However, the energy gap with the 18-meV peak rapidly appears below 40~K, which differs from the temperature dependence of the 15-meV peak at the X point.
In Fig.~\ref{TdepGX}(c2), the temperature dependence of the relative intensity of the 18-meV peak is plotted, and its peak intensity has the same temperature dependence as the magnetic excitation at $Q=(0.5,0.5,0.5)$ by INS~\cite{Alekseev1995}.
The temperature dependence of the 18-meV peak is consistent with that of the magnetic excitation, suggesting that the 18-meV peak at the $\Gamma$ point (i.e., the hybridization band at the $\Gamma$ point) is the initial state of the magnetic excitation.
The appearance of the charge and spin excitations at different $k$-points is consistent with a mean-field theory based on the periodic Anderson model~\cite{Riseborough1992} and with the assumption of the spin exciton theory~\cite{Akbari2009}.
Strictly speaking, however, the peak energies of the magnetic excitations observed by INS and the Raman scattering~\cite{Nyhus1997} are about 14 and 16~meV, respectively, which is slightly lower energy than the 18-meV EDC peak.
This might indicate the property of the spin exciton~\cite{Akbari2009} or there might be magnetic excitations at higher energy region similar to that of other KI, YbB$_{12}$~\cite{Nemkowski2007}.

In the LSDA+$U$ band calculation with Sm$^{3+}$ ions~\cite{Antonov2002}, the energy level of the Sm$^{3+} 5d$ state is close to that of the $4f_{7/2}$ state near the $\Gamma$ point.
Then the hybridization between these states would occur at the $\Gamma$ point.
Due to the opposite logic of the hybridization in the Sm$^{2+}$ ions at the X point, the hybridization in the Sm$^{3+}$ ions means that the mean valence shifts to trivalent.
This is consistent with the evidence of the increase in the mean valence below 30~K~\cite{Mizumaki2009}.

\begin{figure}[t]
\begin{center}
\includegraphics[width=0.45\textwidth]{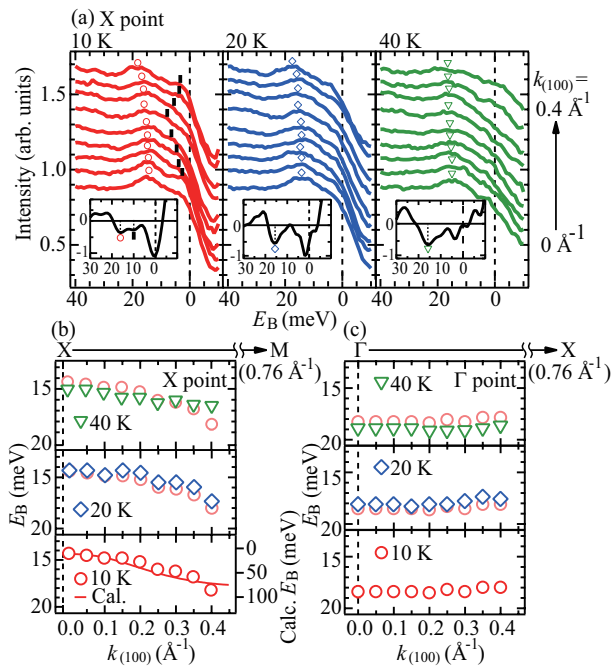}
\end{center}
\caption{
(Color online)
(a) Temperature dependence of energy distribution curves (EDCs) divided by the Fermi-Dirac distribution curve in the X (0~\AA$^{-1}$)--M (0.76~\AA$^{-1}$) line, using 10.6-eV photon energy.
The hybridization band dispersions are shown by open circles, and the observed dispersive in-gap state is shown by vertical lines.
Insets show the second derivative curves of smoothed EDCs at $k_{(100)}=0.25 {\rm \AA}^{-1}$.
The marks correspond to the band dispersions.
Temperature dependence of the hybridization band in the X--M line ($h\nu$~=~10.6~eV, b) and the $\Gamma$--X line ($h\nu$~=~26.0~eV, c).
The band dispersion derived from the LDA calculation (Fig.~\ref{Wide}d1) with the energy scale in the right axis is also plotted by a solid line in (b).
}
\label{TdepX}
\end{figure}
The creation of the hybridization gap is here described in detail.
The temperature-dependent ARPES spectra, divided by the Fermi-Dirac distribution curve along the X--M line, are plotted in Fig.~\ref{TdepX}(a).
At $T$~=~40~K in Fig.~\ref{TdepX}(a), the $k$-dependence of the peak at $E_B$~=~15~meV is almost flat.
The band becomes dispersive below 20~K, as shown in Fig.~\ref{TdepX}(b).
At 10~K, the band dispersion is very similar to that of the LDA band calculation, but the energy of the peak at $k_{(100)}$~=~0.4~\AA$^{-1}$ is about 18~meV, which is much lower than the calculated result of 70~meV.
The width of the experimental $4f$ band dispersion is about 3~meV, which is also much smaller than that of the calculation ($\sim$50~meV).
This suggests that the $4f$ band is strongly renormalized by the strong electron correlation.
Note that the band dispersion near the $\Gamma$ point shown in Fig.~\ref{TdepX}(c) is almost flat, and is also located at much lower energy than that in the LDA calculation ($E_B\sim$ 0.3~eV), as shown in Fig.~\ref{Wide}(d2), indicating a stronger localization at the $\Gamma$ point than that at the X point.
This is consistent with the argument in Fig.~\ref{TdepGX}; i.e., the hybridization gap opens near the X point, while the initial state of the magnetic excitation is the $\Gamma$ point.
The peak energy at the X point is about 15~meV, which is roughly consistent with the valence transition temperature of 120~K ($\sim$10~meV).
However the peak energy of 20~meV at the $\Gamma$ point is much higher energy than the appearance temperature of the magnetic excitation (50~K$~\sim$~4~meV).
This also suggests that the magnetic excitation is strongly correlated.

At $T$~=~10~K, another band dispersion, with an electron-like dispersion at $k\sim$0.25~\AA$^{-1}$ and a hole-like one at $k\sim$0~\AA$^{-1}$, seems to appear at $E_B\sim$8~meV, as shown by vertical lines in Fig.~\ref{TdepX}(a).
The dispersion curve of the in-gap state is very similar to the strong coupled TKI~\cite{Takimoto2011, Dzero2012}.
The detailed discussion is presented in a separated paper~\cite{Kimura2012}.
%
%
%
To summarize, we have investigated the momentum-dependent hybridization state between the conduction band and $4f$ states of SmB$_6$ by using a three-dimensional angle-resolved photoemission spectroscopy.
The temperature dependence of the energy distribution curves suggests that the hybridization state at the binding energy of 15~meV near the X point is the origin of the valence transition below 120~K, while that at 20~meV at the $\Gamma$ point is the initial state of the magnetic excitation at $Q=(0.5,0.5,0.5)$ with strong electron correlation.

We would like to thank Prof. Dzero, Prof. Coleman, and Dr. Matsunami for their fruitful discussion.
Part of this work was performed by the Use-of-UVSOR Facility Program of the Institute for Molecular Science.
The work was partly supported by a Grant-in-Aid for Scientific Research (B) from JSPS of Japan (Grant No.~22340107).
%


\begin{thebibliography}{99}
%
\bibitem{Aeppli1992}
G. Aeppli and Z. Fisk, Comments Condens. Matter Phys. {\bf 16}, No. 3, 155 (1992), and references therein.
%
\bibitem{Travaglini1984}
G. Travaglini and P. Wachter,
Phys. Rev. B {\bf 29}, 893 (1984).
%
\bibitem{Kimura1994}
S. Kimura, T. Nanba, S. Kunii, and T. Kasuya,
Phys. Rev. B {\bf 50}, 1406 (1994).
%
\bibitem{Flachbart2001}
K. Flachbart, K. Gloos, E. Konovalova, Y. Paderno, M. Reiffers, P. Samuely, and P. \v Svec,
Phys. Rev. B {\bf 64}, 085104 (2001).
%
\bibitem{Nozawa2002}
S. Nozawa, T. Tsukamoto, K. Kawai, T. Haruna, S. Shin, and S. Kunii,
J. Phys. Chem. Solids {\bf 63}, 1223 (2002).
%
\bibitem{Souma2002}
S. Souma, H. Kumigashira, T. Ito, T. Takahashi, and S. Kunii,
Physica B {\bf 312-313}, 329 (2002).
%
\bibitem{Cohen1970}
R. L. Cohen, M. Eibsch\"utz, K. W. West, and E. Buehler,
J. Appl. Phys. {\bf 41}, 898 (1970).
%
\bibitem{Tanaka1980}
T. Tanaka, R. Nishitani, C. Oshima, E. Bannai, and S. Kawai,
J. Appl. Phys. {\bf 51}, 3877 (1980).
%
\bibitem{Mizumaki2009}
M. Mizumaki, S. Tsutsui, and F. Iga,
J. Phys.: Conf. Ser. {\bf 176}, 012034 (2009).
%
\bibitem{Kasuya1981}
T. Kasuya, K. Takegahara, Y. Aoki, K. Hanzawa, M. Kasaya, S. Kunii, T. Fujita, N. Sato, H. Kimura, T. Komatsubara, T. Furuno, and J. Rossat-Mignod,
 in Valence Fluctuations in Solids, edited by L. M. Falicov, W. Hanke, and M. B. Maple (North-Holland, Amsterdam, 1981), p. 215.
%
\bibitem{Alekseev1995}
P. A. Alekseev, J. -M. Mignot, J. Rossat-Mignod, V. N. Lazukov, I. P. Sadikov, E. S. Konovalova, and Yu. B. Paderno,
J. Phys.: Condens. Matter {\bf 7}, 289 (1995).
%
\bibitem{Gorshunov1999}
B. Gorshunov, N. Sluchanko, A. Volkov, M. Dressel, G. Knebel, A. Loidl, and S. Kunii,
Phys. Rev. B {\bf 59}, 1808 (1999).
%
\bibitem{Cooley1995}
J. C. Cooley, M. C. Aronson, Z. Fisk, and P. C. Canfield,
Phys. Rev. Lett. {\bf 74}, 1629 (1995).
%
\bibitem{Dzero2010}
M. Dzero, K. Sun, V. Galitski, and P. Coleman,
Phys. Rev. Lett. {\bf 104}, 106408 (2010).
%
\bibitem{Tanaka1975}
T. Tanaka, E. Bannai, S. Kawai, and T. Yamane, 
J. Cryst. Growth {\bf 30}, 193 (1975).
%
\bibitem{Aono1978}
M. Aono, R. Nishitani, T. Tanaka, E. Bannai, and S. Kawai,
Solid State Commun. {\bf 28}, 409 (1978).
%
\bibitem{Ito2007}
T. Ito, S. Kimura, H.J. Im, E. Nakamura, M. Sakai, T. Horigome, K. Soda, and T. Takeuchi,
AIP Conf. Proc. {\bf 879}, 587 (2007).
%
\bibitem{Kimura2010}
S. Kimura, T. Ito, M. Sakai, E. Nakamura, N. Kondo, K. Hayashi, T. Horigome, M. Hosaka, M. Katoh, T. Goto, T. Ejima, and K. Soda,
Rev. Sci. Instrum. {\bf 81}, 053104 (2010). 
%
\bibitem{WIEN2k}
P. Blaha
, K. Schwarz, P. Sorantin and S. B. Trickey, 
Comput. Phys. Commun. {\bf 59}, 399 (1990).
%
\bibitem{Denlinger2000}
J. D. Denlinger, G. -H. Gweon, J. W. Allen, C. G. Olson, Y. Dalichaouch, B. -W. Lee, M. B. Maple, Z. Fisk, P. C. Canfield, P. E. Armstrong, 
Physica B {\bf 281\&282}, 716 (2000).
%
\bibitem{Yeh1985}
J. J. Yeh and I. Lindau,
Atomic Data and Nuclear Data Tables {\bf 32}, 1 (1985).
%
\bibitem{Riseborough1992}
P. S. Riseborough,
Phys. Rev. B {\bf 45}, 13984 (1992).
%
\bibitem{Akbari2009}
A. Akbari, P. Thalmeier, and P. Fulde,
Phys. Rev. Lett. {\bf 102}, 106402 (2009).
%
\bibitem{Nyhus1997}
P. Nyhus, S. L. Cooper, Z. Fisk, and J. Sarrao,
Phys. Rev. B {\bf 55}, 12488 (1997).
%
\bibitem{Antonov2002}
V. N. Antonov, B. N. Harmon, and A. N. Yaresko,
Phys. Rev. B {\bf 66}, 165209 (2002).
%
\bibitem{Nemkowski2007}
K. S. Nemkovski, J.-M. Mignot, P. A. Alekseev, A. S. Ivanov, E. V. Nefeodova, A. V. Rybina, L.-P. Regnault, F. Iga, and T. Takabatake,
Phys. Rev. Lett. {\bf 99}, 137204 (2007).
%
\bibitem{Takimoto2011}
T. Takimoto,
J. Phys. Soc. Jpn. {\bf 80}, 123710 (2011).
%
\bibitem{Dzero2012}
M. Dzero, K. Sun, P. Coleman, and V. Galitski,
Phys. Rev. B {\bf 85}, 045130 (2012).
%
\bibitem{Kimura2012}
S. Kimura, T. Hajiri, M. Matsunami, H. Miyazaki, and S. Kumii,
unpublished data.
%
\end{thebibliography}
\end{document}